\documentclass[useAMS,twocolumn,usenatbib]{mn2e}

\usepackage{graphicx}

\title[Spherically symmetric black hole accretion 
in Schwarzschild geometry]
{Critical properties of spherically symmetric black hole accretion  
in Schwarzschild geometry}
\author[Mandal et al.]
{Ipsita Mandal,$^{1}$\thanks{ipsita@mri.ernet.in}
Arnab K. Ray$^{2}$\thanks{akr@iucaa.ernet.in}
and Tapas K. Das$^{1}$\thanks{tapas@mri.ernet.in}\\
$^{1}$Harish--Chandra Research Institute, Chhatnag Road, Jhunsi,
Allahabad 211019, India\\
$^{2}$Inter--University Centre for Astronomy and Astrophysics, Post
Bag 4, Ganeshkhind, Pune University Campus, Pune 411007, India}

\begin{document}


\maketitle

\label{firstpage}

\begin{abstract}
The stationary spherically symmetric accretion flow in the Schwarzschild
metric has been set up as an autonomous first-order dynamical system, 
and it has been studied completely analytically. Of the 
three possible critical points in the flow, the one that is physically 
realistic behaves like the saddle point of the standard Bondi accretion
problem. One of the two remaining critical points 
exhibits the strange mathematical behaviour of being either a saddle 
point or a centre-type point, depending on the values of the flow 
parameters. The third critical point is always unphysical and behaves
like a centre-type point. The treatment has been extended to 
pseudo-Schwarzschild flows for comparison with the general relativistic 
analysis. 
\end{abstract}


\section{Introduction}
\label{sec1}

To researchers in astrophysics and general relativity,
physical models of spherical symmetry have an abiding appeal. One 
especial advantage with these models is that almost always they lend
themselves to an exact mathematical analysis, and in the process they
allow a very clear insight to be had into the underlying physical 
principles. For this reason in particular, spherically symmetric 
models frequently serve as a firm foundation for constructing theoretical
models of physical systems involving more realistic and, therefore,  
unavoidably complicated features. 

Studies in accretion are no exception to this practice. Ever since
the seminal work published by~\citet{bon52}, that effectively 
launched the subject in the form in which it is recognised today,
the problem of spherically symmetric flows has been revisited time
and again from various angles~\citep{par58,par66,an67,bal72,mich72,
mes75,blum76,ms77,beg78,cos78,sb78,gar79,bri80,monc80,pso80,vit84,
bona87a,bona92b,td92,km94,mar95,tuf95,tmk96,zmt96,tmk97,ke98,das99,
malec99,ttsrcl99,das00,ds01,rb02,ray03,das04,rb05,gai06}. 

The original work of~\citet{bon52} introduced formal fluid 
dynamical equations in the Newtonian construct of space and time to 
study the stationary accretion problem. From here the connection 
to the proper general relativistic framework was not too long in 
coming. In particular it was~\citet{mich72} who made an important 
early foray into the general relativistic domain, which was 
followed by a spate of other works, of which some (to mention a few),
along with the paper of~\citet{mich72},  
addressed various aspects related to the critical behaviour of 
general relativistic flows in spherical symmetry~\citep{beg78,bri80,
malec99,ds01}. 

The work being presented here is also along the same lines. Its
direct purpose is to construct a pedagogical theory to understand 
the nature of the critical 
points of stationary spherically symmetric flows in the Schwarzschild 
metric, after starting with the set of basic stationary equations which 
govern the flow. To the extent
that critical solutions --- specifically transonic solutions in 
regard to spherically symmetric flows --- can only 
pass through some critical points (which must arguably be saddle points), 
this treatment will also have a bearing on a very important general 
issue in accretion studies --- the manner in which a compressible 
astrophysical fluid passes (either continuously or discontinuously) 
from infinity to the event horizon of a black hole. 
In addressing this question the mathematical method that has been 
adopted is a dynamical systems analysis, which is always an 
effective tool for researchers in non-linear dynamics, and which has 
been tried successfully before in a related astrophysical
system, namely multitransonic flows in an inviscid and thin 
pseudo-Schwarzschild accretion disc~\citep{crd06}. While it is 
to be naturally expected that the same treatment could be carried 
over directly to a rotational flow in the Schwarzschild metric, it
has also been a worthwhile exercise to consider spherically symmetric
flows first, as a suitably uncomplicated starting point into more 
intricate general relativistic problems. An immediate advantage in 
studying this relatively simple system has been that the mathematical
treatment could be carried out with full analytic rigour, something,
as it has been stressed right at the beginning, that has inspired 
the choice of the spherically symmetric model in the first place. 
It has been found here that under practical restrictions on the 
conditions for critical behaviour, it will be possible to gain 
a complete mathematical understanding (in the sense of producing 
final results which are absolutely non-numerical) of the nature 
of the critical points (not all of whom may be physically realistic),
and the pattern of the solution topologies in the neighbourhood of
those points. In no way do these results contradict any of the earlier
findings~\citep{mich72,beg78}, and if anything, many surprisingly 
new features in the flow, hitherto unrecognised, have been revealed. 

The dynamical systems treatment has also been carried out on  
pseudo-Schwarzschild spherical flows driven by some of the established
pseudo-Newtonian potentials~\citep{pw80,nw91,abn96} to check for the
consistency of this approach with the fully general relativistic methods. 
All the results have been in complete qualitative compatibility with 
one another in the sense that there is only one physically meaningful
critical point in the flow through which a solution could pass 
transonically, connecting infinity to the event horizon of the 
black hole. In one important detail, however, a quantitative difference 
has appeared. For accretion governed by cold ambient conditions, 
transonicity has been shown to be very much more pronounced in a 
properly relativistic flow, than in a pseudo-Schwarzschild flow. 

The case for a dynamical systems approach in studying a proper general 
relativistic problem has been argued cogently in this work. The 
mathematical methods demonstrated here can very well be applied to
the more involved accretion disc system, described both by the 
Schwarzschild metric and the Kerr metric. This particular
study will be reported separately. Meanwhile, following the treatment
on pseudo-Schwarzschild flows reported by~\citet{crd06}, the present 
work on spherically symmetric flows may be considered to be the second 
in a series that pedagogically underlines the conspicuous advantages 
of applying dynamical systems methods in standard astrophysical fluid
flow problems. 

\section{The general relativistic flow and its fixed points}
\label{sec2}

In this general relativistic treatment of a spherically symmetric, 
stationary, compressible fluid flow, the two relevant flow variables 
will be the radial inflow three-velocity, $v$, and the local proper 
mass density of the fluid, $\rho$. The radial coordinate 
of the flow is scaled by the Schwarzschild radius, 
$r_{\mathrm{g}} = 2GM_{\mathrm{BH}}/c^2$ (here $M_{\mathrm{BH}}$ is 
the mass of the black hole), with any characteristic velocity in the 
flow being scaled by $c$. Setting $G = c = 1$, the 
general relativistic analogue of Bernoulli's equation will be given 
as~\citep{skc90,skc96,das04} 
\begin{equation}
\label{bernou1}
{\mathcal E} = h v_{t} = \frac{p + \epsilon}{\rho} 
\sqrt{\frac{1 - r^{-1}}{1 - v^2}} , 
\end{equation}
with the pressure, $p$, connected to the density, $\rho$, through an 
equation of state, $p = k \rho^{\gamma}$, and the specific enthalpy,
$h$, expressed as 
\begin{equation}
\label{enthal}
h = \frac{p + \epsilon}{\rho} , 
\end{equation}
in which the energy density, $\epsilon$ (which includes the rest mass
density and the internal energy), is to be set down as, 
\begin{equation}
\label{epsi}
\epsilon = \rho + \frac{p}{\gamma - 1} . 
\end{equation}
It is then possible to arrive at the relation 
\begin{equation}
\label{bernou2}
{\mathcal E} = \left(\frac{\gamma -1}{\gamma -1 - c_{\mathrm s}^2}\right)
\sqrt{\frac{1 - r^{-1}}{1 - v^2}} , 
\end{equation}
with the speed of sound, $c_{\mathrm s}$, defined under conditions of 
constant entropy, $\mathcal S$, as 
\begin{equation}
\label{acous}
c_{\mathrm s}^2 =\frac{\partial p}{\partial \epsilon}\bigg\vert_{\mathcal S} .
\end{equation}
Through the equation state, $p = k \rho^{\gamma}$, the speed of sound can 
be connected to the density, $\rho$, as 
\begin{equation}
\label{conden}
\rho = \left[\frac{c_{\mathrm s}^2}{\gamma k
\left(1 - n c_{\mathrm s}^2\right)}\right]^n , 
\end{equation}
with $n$ being given by the usual definition of the polytropic 
index~\citep{sc39} as $n = (\gamma -1 )^{-1}$. 

The stationary continuity condition, on the other hand, will give 
another relation connecting the velocity and density fields as, 
\begin{equation}
\label{cont}
4 \pi \rho v r^2 \sqrt{\frac{1 - r^{-1}}{1 - v^2}} = \dot{m} , 
\end{equation}
with $\dot{m}$ being an integration constant, which is to be physically
identified as the mass accretion rate. 

It is now easy to see that equations~(\ref{bernou2}),~(\ref{conden})
and~(\ref{cont}) will give a complete description of the flow system. 
Making use of equation~(\ref{conden}) in equation~(\ref{cont}), and 
then going back to equation~(\ref{bernou2}), it will be possible to 
express the gradient of solutions in the $r$ --- $v^2$ plane as 
\begin{equation}
\label{dvdr}
\frac{\mathrm d}{{\mathrm d} r}(v^2) = 
\frac{v^2\left(1-v^2\right)\left[c_{\mathrm s}^2\left(4r-3\right)-1\right]}
{r \left(r-1\right) \left(v^2-c_{\mathrm s}^2\right)} , 
\end{equation}
with $c_{\mathrm s}$ being used as a characteristic local scale of velocity
in the fluid, against which the local
bulk velocity of the flow, $v$, is to be measured. 

From the foregoing expression it is evident that there will be non-trivial
singularities under the conditions $r=1$ and $v^2=c_{\mathrm s}^2$, 
unless the numerator in equation~(\ref{dvdr}) vanishes simultaneously. 
The condition $r=1$, of course, corresponds to the behaviour of the 
flow on the actual event horizon of the black hole (where $v^2=1$), 
but of immediate interest is the condition $v^2=c_{\mathrm s}^2$, 
which will give a critical condition for the flow at $r>1$, if and 
only if the requirement of $c_{\mathrm s}^2\left(4r-3\right)=1$, is 
simultaneously satisfied in the numerator. In that event the critical
point conditions in the flow will be expressed as
\begin{equation}
\label{critcon}
v_{\mathrm c}^2 = c_{\mathrm{sc}}^2 = \frac{1}{4r_{\mathrm c} - 3} ,
\end{equation}
with the subscript ``$\mathrm c$" labelling the critical point values. 

The next logical step from here is to represent the critical point 
coordinates in terms of the parameters of the system. This can be 
done in two ways --- either by substituting $v_{\mathrm c}$ and 
$c_{\mathrm{sc}}$ in equation~(\ref{bernou2}) and expressing 
$r_{\mathrm c}$ as a function of $\mathcal E$ and $\gamma$, or by
substituting the same critical point values in equation~(\ref{cont}) 
and expressing $r_{\mathrm c}$ as a function of $\dot{m}$ and $\gamma$. 
Either approach is entirely equivalent to the other, and here for 
simplicity of algebraic manipulations, the former approach is being
adopted. This will deliver a cubic equation in $r_{\mathrm c}$ as
\begin{equation}
\label{cubic}
r_{\mathrm c}^3 + {\mathcal A}_2 r_{\mathrm c}^2 + {\mathcal A}_1 
r_{\mathrm c} + {\mathcal A}_0 = 0 , 
\end{equation}
where 
\begin{displaymath}
\label{a0}
{\mathcal A}_0=\frac{27}{64\left({\mathcal E}^2-1\right)} ,
\end{displaymath}
\begin{displaymath}
\label{a1}
{\mathcal A}_1=\frac{\left(2-3\gamma\right)^2{\mathcal E}^2 
- 27 \left(\gamma -1 \right)^2}{16 \left({\mathcal E}^2 - 1 \right)
\left(\gamma -1 \right)^2} 
\end{displaymath}
and 
\begin{displaymath}
\label{a2}
{\mathcal A}_2=\frac{2\left(2-3\gamma\right){\mathcal E}^2 +  
9\left(\gamma -1 \right)}{4\left({\mathcal E}^2 - 1 \right)
\left(\gamma -1 \right)} . 
\end{displaymath}

The solutions of equation~(\ref{cubic}) can be found completely 
analytically by employing the Cardano-Tartaglia-del Ferro method
for solving cubic equations. To that end it should be first convenient
to define 
\begin{displaymath}
\label{sigma1}
\Sigma_1 = \frac{3{\mathcal A}_1 -{\mathcal A}_2^2}{9} , 
\end{displaymath}
\begin{displaymath}
\label{sigma2}
\Sigma_2 = \frac{9{\mathcal A}_1 {\mathcal A}_2 - 27 {\mathcal A}_0
- 2 {\mathcal A}_2^3}{54} , 
\end{displaymath}
\begin{displaymath}
\label{psi}
\Psi = \Sigma_1^3 + \Sigma_2^2 , 
\end{displaymath}
\begin{displaymath}
\label{xi1}
\xi_1 = \left(\Sigma_2 + \sqrt{\Psi} \right)^{1/3}
\end{displaymath}
and 
\begin{displaymath}
\label{xi2}
\xi_2 = \left(\Sigma_2 - \sqrt{\Psi} \right)^{1/3} , 
\end{displaymath}
following which, the three roots of $r_{\mathrm c}$ can 
ultimately be set down as
\begin{equation}
\label{root1}
r_{\mathrm{c1}} = - \frac{{\mathcal A}_2}{3} + \left(\xi_1 + \xi_2\right) , 
\end{equation} 
\begin{equation}
\label{root2}
r_{\mathrm{c2}} = - \frac{{\mathcal A}_2}{3} - \left(\xi_1 + \xi_2\right)
+ {\mathrm i}\frac{{\sqrt{3}}}{2} \left(\xi_1 - \xi_2\right)
\end{equation}
and
\begin{equation}
\label{root3}
r_{\mathrm{c3}} = - \frac{{\mathcal A}_2}{3} - \left(\xi_1 + \xi_2\right)
- {\mathrm i}\frac{{\sqrt{3}}}{2} \left(\xi_1 - \xi_2\right) , 
\end{equation}
respectively. 

The sign of $\Psi$ should be crucial in determining the nature of the 
roots. If $\Psi >0$, then only one root of $r_{\mathrm c}$ will be
real, while for $\Psi <0$, there will be three real roots. Since it is
obvious that $\Psi$ will have a dependence on $\mathcal E$ and $\gamma$,
it will be instructive to consider the appropriate ranges of values for
these two parameters. 

The parameter $\mathcal E$ is scaled in terms of the rest mass energy
and it includes the rest mass energy itself. So it might be argued that
a lower limit
of $\mathcal E$ would be $\mathcal E =1$. On the other hand, although 
$\mathcal E$ can in principle assume any value greater than unity, values
of $\mathcal E >2$ will imply extremely hot conditions at the outer 
boundary, with the thermal energy being much greater than the rest mass
energy. Such a situation could not conceivably prevail in realistic
astrophysical systems, and so the practically admissible range of 
$\mathcal E$ will be restricted to $1<{\mathcal E}<2$. 
The non-relativistic range of $\mathcal E$, on the other hand, will be 
$0<{\mathcal E}<1$, without any involvement of the rest mass energy. 
This range will be considered for the pseudo-Schwarzschild treatment in 
Section~\ref{sec4}. 

The parameter $\gamma$ is likewise restricted by the range $1<\gamma <2$. 
The lower limit, i.e. $\gamma =1$, corresponds to optically thin, 
isothermal accretion, while values of $\gamma >2$ will involve 
magneto-hydrodynamics in the general relativistic theory and the
anisotropic nature of pressure. For most realistic purposes, it should 
be noted, the range of $\gamma$ actually varies from $4/3$ to $5/3$. 
Detailed discussions devoted to these issues are to be found in the 
literature~\citep{beg78,das04}. 

And so it is that with the ranges of $1< {\mathcal E} <2$ and 
$1< \gamma <2$, it should be easy to show that $\Psi$ would always
be negative. This will consequently imply that the three roots of 
$r_{\mathrm c}$, as given by equations~(\ref{root1}),~(\ref{root2})
and~(\ref{root3}), are always real, and they can be represented in 
terms of a new variable, 
\begin{displaymath}
\label{theta}
\Theta = \arccos \left(\frac{\Sigma_2}{\sqrt{- \Sigma_1^3}} \right) , 
\end{displaymath}
as
\begin{equation}
\label{rootfin}
r_{{\mathrm c}j} = - \frac{{\mathcal A}_2}{3} + 2 \sqrt{- \Sigma_1} 
\cos\left[\frac{\Theta + 2\pi \left(j -1 \right)}{3} \right] , 
\end{equation}
with the label $j$ taking the values $\{j=1,2,3\}$, respectively, 
for the three distinct roots. Of these roots, $r_{\mathrm{c2}}$ is 
always negative
and, therefore, is not of much physical interest. The other two roots
are always positive, and of these two, the one at $r_{\mathrm{c1}}$ 
is always to be found at distances greater than 
the event horizon of the black hole, i.e. $r_{\mathrm{c1}} >1$. 
A physically 
meaningful transonic inflow solution, connecting infinity to the 
event horizon, seems to prefer the critical point at $r_{\mathrm{c1}}$
to the one at $r_{\mathrm{c3}}$, even when $r_{\mathrm{c3}} >1$. 
Through the latter point, the flow 
exhibits non-physical properties like the matter inflow rate, as 
given by equations~(\ref{conden}) and~(\ref{cont}), being reduced 
to an imaginary quantity. It was also pointed out by~\citet{das04}
that a flow associated with this point becomes superluminal much 
before reaching the event horizon of the black hole. Nevertheless, for
all its apparent barrenness from a physical perspective, this critical
point is not entirely devoid of some very interesting mathematical
properties, when a dynamical systems approach is made to study the 
nature of the critical points of the flow. This issue will be taken 
up in Section~\ref{sec3}. 

\section{Properties of the fixed points : An autonomous dynamical system}
\label{sec3}

So far the flow variables have been ascertained only at the critical
points. Since the flow equations are in general non-linear differential
equations, short of carrying out a numerical integration, there is no
completely rigorous analytical prescription for solving these differential
equations to determine the global nature of the flow variables. 
Nevertheless, some analytical headway could be made after all by taking
advantage of the fact that equation~(\ref{dvdr}), which gives a 
complete description of the $r$ --- $v^2$ phase portrait of the flow,
is an autonomous first-order differential equation, and as such, could
easily be recast into the mathematical form ${\dot{x}} = X(x,y)$ and
${\dot{y}} = Y(x,y)$, which is that of the very familiar coupled 
first-order dynamical system~\citep{js99}. 

With the adoption of this line of attack, equation~(\ref{dvdr}) may be
decomposed in terms of a mathematical parameter, $\tau$, to read as
\begin{eqnarray}
\label{dynsys}
\frac{\mathrm{d}}{\mathrm{d}\tau}(v^2) &=& v^2 \left(1-v^2\right)
\left[c_{\mathrm s}^2\left(4r-3\right)-1\right] \nonumber \\
\frac{\mathrm{d}r}{\mathrm{d} \tau} &=& r 
\left(r-1\right) \left(v^2-c_{\mathrm s}^2\right)
\end{eqnarray}
in both of which, it must be noted that the parameter $\tau$ does not
make an explicit appearance in the right hand side, something of an 
especial advantage that derives from working with autonomous systems. 
This kind of parametrization is quite common in fluid dynamics~\citep{bdp93},
and in accretion studies especially, this approach has been made 
before~\citep{rb02,ap03,crd06}. A further point that has to be noted
is that the function $c_{\mathrm s}^2$ in the right hand side of 
equations~(\ref{dynsys}) can be expressed entirely in terms of $v^2$ 
and $r$, with the help of equations~(\ref{conden}) and~(\ref{cont}).
This will exactly satisfy the criterion of a first-order dynamical system. 

The critical points of the foregoing dynamical system, as 
equations~(\ref{critcon}) give them, have already been identified,
and as equation~(\ref{cubic}) indicates, they have also been fixed
in terms of the physical flow parameters. Beyond this stage, the 
next task would be to make a linearised approximation about the 
fixed point coordinates and extract a linear dynamical system out
of equations~(\ref{dynsys}). This will give a direct way to establish
the nature of the critical points (or fixed points), which will
ultimately pave the way for an investigation into the global behaviour
of the solutions in the phase portrait of the flow. Indeed, not 
infrequently, if the flow system is simple enough, with only an 
understanding of the features of its critical points, complete 
qualitative predictions can be made about the global solutions. 
In this regard the classical spherically symmetric Bondi flow,
with its single critical point, provides an object lesson~\citep{rb02}. 

Expanding about the fixed point values, a perturbation scheme of 
the kind $v^2 = v_{\mathrm{c}}^2 + \delta v^2$, $c_{\mathrm{s}}^2 =
c_{\mathrm{sc}}^2 + \delta c_{\mathrm{s}}^2$ and
$r = r_{\mathrm{c}} + \delta r$ is now to be applied  
on equations~(\ref{dynsys}), and then 
a set of coupled autonomous linear equations is to be derived from
them. While doing so, it will also be necessary to express 
$\delta c_{\mathrm{s}}^2$ itself in terms of $\delta r$ and $\delta v^2$, 
with the help of equations~(\ref{conden}) and~(\ref{cont}), as
\begin{equation}
\label{varsound}
\frac{\delta c_{\mathrm{s}}^2}{c_{\mathrm{sc}}^2} = 
-\frac{\gamma-1-c_{\mathrm{sc}}^2}{2}\left[\frac{1}{1 - v_{\mathrm{c}}^2}
\left(\frac{\delta v^2}{v_{\mathrm{c}}^2}\right)
+ \frac{4r_{\mathrm{c}} -3}{r_{\mathrm{c}}-1}
\left(\frac{\delta r}{r_{\mathrm{c}}}\right)\right] . 
\end{equation}
The resulting coupled linearised equations will then read as 
\begin{eqnarray}
\label{lindyn}
\frac{\mathrm{d}}{\mathrm{d}\tau}(\delta v^2) &=&  
- {\mathcal B}\, \delta v^2 + \left[4c_{\mathrm{sc}}^2 - 
\frac{{\mathcal B}\left(1-c_{\mathrm{sc}}^2\right)}
{r_{\mathrm{c}}\left(r_{\mathrm{c}}-1\right)}\right]\delta r \nonumber \\
\frac{\mathrm{d}}{\mathrm{d}\tau}(\delta r)
&=& r_{\mathrm{c}}\left(r_{\mathrm{c}} -1 \right)
\left(1 + \frac{{\mathcal B}}{1 - v_{\mathrm{c}}^2}\right) \delta v^2 
+ {\mathcal B}\, \delta r , 
\end{eqnarray}
with ${\mathcal B} = (\gamma - 1 - c_{\mathrm{sc}}^2)/2$. Using 
solutions of the type $\delta v^2 \sim \exp(\Omega \tau)$
and $\delta r \sim \exp(\Omega \tau)$ in equations~(\ref{lindyn}),
the eigenvalues of the stability matrix associated with the critical
points will be derived as 
\begin{equation}
\label{eigen}
\Omega^2 = \frac{{\mathcal C}^2}{4} + \left(2r_{\mathrm{c}} 
- {\mathcal C}\right)\left[\frac{3\left(r_{\mathrm{c}}-1\right)}
{4r_{\mathrm{c}}-3} - \frac{2 - \gamma}{4} \right] , 
\end{equation}
where 
\begin{displaymath}
{\mathcal C} = \frac{4\left(\gamma -1 \right)r_{\mathrm{c}}
- \left(3 \gamma -2 \right)}{4r_{\mathrm{c}} -3} . 
\end{displaymath}

Once the position of a critical point, $r_{\mathrm{c}}$, has become
known, it is then quite easy to determine the nature of that critical 
point by using $r_{\mathrm{c}}$ in equation~(\ref{eigen}).
Since $r_{\mathrm{c}}$ is a function of $\mathcal E$ and $\gamma$,  
it effectively implies that $\Omega^2$ can, in principle, be
regarded as a function of the flow parameters.
From the form of $\Omega^2$ in equation~(\ref{eigen}), a generic 
conclusion that can be immediately drawn is that the only possible 
critical points will be saddle points and centre-type points, and 
for the former, $\Omega^2 >0$, while for the latter, $\Omega^2 <0$.
This is entirely to be expected because the physical system under
study here is a conservative system, very much like, by analogy, 
the undamped simple harmonic oscillator, the fixed points of whose 
phase portrait also manifest identical properties~\citep{js99}. 

Of the three critical points, as implied by equation~(\ref{rootfin}),
the one given by the unphysical negative root, $r_{\mathrm{c2}}$, is
always a centre-type point. This is something that by itself is also 
an equally unphysical trait as far as transonic accretion is concerned,
where the whole objective is to have a solution that
will connect infinity to the event horizon of the black hole,
and in doing so will cross the sonic barrier with a finite gradient. 

\begin{figure}
\begin{center}
\includegraphics[scale=0.33, angle=-90]{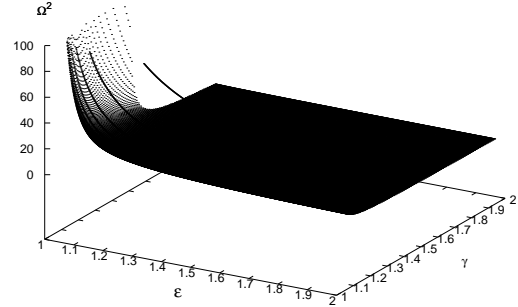}
\caption{\label{f1} \small{Variation of $\Omega^2$ (associated with 
a physical saddle point) with respect to the parameters $\mathcal E$ 
and $\gamma$ for the fully general relativistic spherically symmetric
flow. All values of $\Omega^2$ are positive for the chosen ranges of 
$\mathcal E$ and $\gamma$. For small values of these two parameters,
the saddle-type feature is very robust.}}
\end{center}
\end{figure}

\begin{figure}
\begin{center}
\includegraphics[scale=0.33, angle=-90]{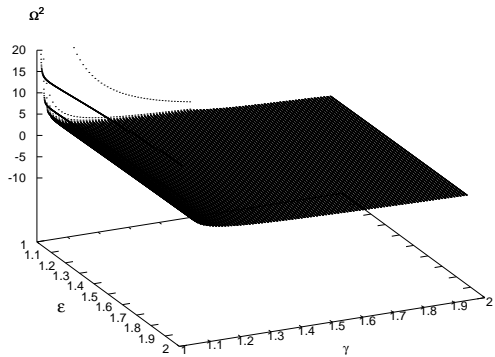}
\caption{\label{f2} \small{The general relativistic flow has a strange
critical point which behaves sometimes like a saddle $(\Omega^2 >0)$ and 
at other times like a centre-type point $(\Omega^2 <0)$, depending on the 
values of the parameters $\mathcal E$ and $\gamma$. The dependence of 
$\Omega^2$ does not exhibit widespread deviations, except for small 
values of $\mathcal E$ and $\gamma$.}}
\end{center}
\end{figure}

This requirement, on the other hand, is very eminently met at the 
critical point fixed at $r_{\mathrm{c1}}$, which is always a saddle 
point and allows a physical transonic solution to pass throught itself
without any hindrance whatsoever. The behaviour of this critical point
has been graphically depicted in Fig.~\ref{f1}. An interesting fact 
that emerges from the plot is that for low values of $\mathcal E$ and 
$\gamma$, there is a strong growth behaviour for $\Omega^2$. In fact, 
as $\mathcal E$ and $\gamma$ approach unity, the value of $\Omega^2$ 
increases by four orders of magnitude than what has been scaled along
the vertical axis of Fig.~\ref{f1}\footnote{This reduced scaling of
the vertical axis will give a much better resolution of the overall 
behaviour of $\Omega^2$ against the parameters on which it is dependent.}, 
and this is exactly how it should be. Apart from the sign of $\Omega^2$ 
--- always positive and, therefore, indicative of a saddle point --- its
magnitude also conveys quantitative information about the strength of 
the saddle-type behaviour; the greater its magnitude, the more prominent
in effect will be the transonic behaviour of the flow (which will pass
through the saddle point). When $\mathcal E$ and $\gamma$ approach unity,
it will correspond more closely to a cold, isothermal distribution of
matter, and this will make the flow easily submit to the strong 
gravitational influence of the black hole (a hotter and more 
pressure-dominated flow will be capable of building a much greater
resistance against gravity). Transonicity can only occur when gravity
triumphs over all other effects, and the general features of the flow
indicated by Fig.~\ref{f1} is an emphatic endorsement of physical 
transonicity in a general relativistic scenario.  

On the other hand, a most intriguing and counter-intuitive behaviour is 
to be encountered at the critical 
point characterised by $r_{\mathrm{c3}}$, which is placed between 
$r_{\mathrm{c1}}$ (always a saddle point) and $r_{\mathrm{c2}}$ 
(always a centre-type point). Depending on the values of 
the parameters $\mathcal E$ and $\gamma$, this point exhibits the 
properties of both a saddle point ($\Omega^2 >0$) and a centre-type
point ($\Omega^2 <0$). As far as the former case is concerned, this is 
a very curious state of affairs indeed, because conventional wisdom about 
well-behaved fixed points will have it that no two adjacent fixed points 
can both be saddle points~\citep{js99},
quite contrary to what is being seen here --- a saddle point (obviously
realistic and physically meaningful) and a centre-type point (however
physically unrealistic) flanking a point which, on some occasions at 
least, behaves like a saddle (and on other occasions like a centre). 
It is difficult to provide an analogy for this kind of behaviour from
any other area in physics. One might conjecture that this waywardness
could be intimately connected to the divergent behaviour (like 
superluminal motion) exhibited by solutions associated with this
critical point (when it behaves like a saddle point). A quantitative 
graphical understanding of the nature
of this critical point has been conveyed in Fig.~\ref{f2}. Once again 
it is to be seen that as $\mathcal E$ and $\gamma$ both assume values 
closer to unity, strong evidence of a saddle-type behaviour results. 

\section{The pseudo-Schwarzschild approach : A comparative study} 
\label{sec4}

Frequently in studies of black hole accretion, it becomes convenient
to dispense completely with the rigour of general relativity, and instead
make use of an ``effective" pseudo-Newtonian potential that will imitate
general relativistic effects in the Newtonian construct of space and
time. In that event the relevant stationary equations for the compressible
spherically symmetric flow will look like
\begin{equation}
\label{euler}
v \frac{\mathrm{d}v}{\mathrm{d}r}
+ \frac{1}{\rho}\frac{\mathrm{d}p}{\mathrm{d}r}
+ \phi^{\prime}(r) = 0
\end{equation}
and 
\begin{equation}
\label{con}
\frac{\mathrm{d}}{\mathrm{d}r}\left(\rho vr^2 \right) = 0
\end{equation}
respectively, with the former being the familiar Euler's equation and 
the latter the equation of continuity~\citep{skc90,fkr02}. In 
equation~(\ref{euler}), $\phi(r)$ is the generalised pseudo-Newtonian 
potential driving the flow (with the prime denoting its spatial derivative),
and $p$ is the pressure of the flowing gas, which is related to the 
density by the usual polytropic prescription. The local speed of sound, 
with which the bulk flow will have to be scaled, is defined by
$c_{\mathrm s}^2 = \partial p/\partial \rho$, following which the 
connection between the $\rho$ and $c_{\mathrm s}$ could be established as
\begin{equation}
\label{sound}
\rho = \left(\frac{c_{\mathrm s}^2}{\gamma k}\right)^n ,
\end{equation}
whose form may, for academic interest, be compared with the general 
relativistic analogue given in equation~(\ref{conden}). 

Making use of equation~(\ref{sound}) in equation~(\ref{con}), and then
going back to equation~(\ref{euler}), will lead to a relation for the 
gradient of solutions, which will read as 
\begin{equation}
\label{dvdr1}
\frac{\mathrm d}{{\mathrm d} r}(v^2) =
\frac{2v^2\left[2c_{\mathrm s}^2-r\phi^{\prime}(r)\right]}
{r \left(v^2-c_{\mathrm s}^2\right)} .
\end{equation}

The critical points in the flow will be derived from the standard
requirement that the flow solutions will have a finite gradient when
they will cross the sonic horizon, which will mean that both the 
numerator and the denonimator will have to vanish simultaneously 
and non-trivially. This can only happen when 
\begin{equation}
\label{critcon1}
v_{\mathrm c}^2 = c_{\mathrm{sc}}^2 = 
\frac{r_{\mathrm c}\phi^{\prime}(r_{\mathrm c})}{2} ,
\end{equation}
which gives the critical point conditions, 
with the subscript ``$\mathrm c$" labelling the critical point values,
as usual.

It is not a difficult exercise to integrate equation~(\ref{euler})
and then transform the variable $\rho$ in it to $c_{\mathrm s}$, with 
the help of equation~(\ref{sound}). Once this has been done, the 
critical conditions, as given by equations~(\ref{critcon1}), will
have to be invoked, and all of these will deliver a relation for
fixing the critical point coordinates in terms of the flow parameters,
$\mathcal E$ (which is actually Bernoulli's constant) and $\gamma$. 
This relation will look like 
\begin{equation}
\label{eulerfix}
\frac{1}{4} \left(\frac{\gamma +1}{\gamma -1}\right) 
r_{\mathrm c}\phi^{\prime}(r_{\mathrm c}) + \phi(r_{\mathrm c})
= {\mathcal E}, 
\end{equation}
with the ranges of values of $\mathcal E$ and $\gamma$ here, in 
what is essentially a non-relativistic approach, being accordingly 
chosen~\citep{ds01}, as opposed to the relativistic values of
$\mathcal E$ and $\gamma$ adopted in Section~\ref{sec3}. 

The choice of the pseudo-Newtonian potential, $\phi(r)$, will 
obviously determine the number of roots of equation~(\ref{eulerfix}).  
Four such potentials have been considered here, and they have in 
general been labelled as $\phi \equiv \phi_i (r)$, with
$\{i=1,2,3,4\}$. In an explicit form, each of these potentials will 
be given as 
\begin{eqnarray}
\label{potens}
\phi_1 (r) &=& - \frac{1}{2 \left(r - 1 \right)} \nonumber \\
\phi_2 (r) &=& - \frac{1}{2r} \left[ 1 - \frac{3}{2r} + 12
\left( \frac{1}{2r} \right)^2 \right ] \nonumber \\
\phi_3 (r) &=& -1 + \left(1 - \frac{1}{r} \right)^{1/2} \nonumber \\
\phi_4 (r) &=& \frac{1}{2} \ln \left(1 - \frac{1}{r} \right)
\end{eqnarray}
in all of which, the length of the radial coordinate, $r$, has been
scaled in units of the Schwarzschild radius, defined as
$r_{\mathrm g} = 2GM_{\mathrm{BH}}/c^2$. Every potential mentioned above 
has been introduced in accretion literature at various stages to meet 
some specific physical requirement --- 
$\phi_1$ by~\citet{pw80}, $\phi_2$ by~\citet{nw91}, and $\phi_3$ and
$\phi_4$ by~\citet{abn96}, respectively. With respect to spherically 
symmetric flows in particular, a comparative overview of the physical 
properties of these potentials has been given by~\citet{ds01}. 

Considering each of the potentials separately in equation~(\ref{eulerfix}),
it will be seen that two distinct roots will be obtained on using both 
$\phi_1$ and $\phi_3$, while from $\phi_2$ three roots will be delivered. 
The fourth potential, $\phi_4$, will lead to a transcendental equation,
and any root, therefore, can only be extracted by numerical methods. 
Using the bisection algorithm, it can be shown that only one physical 
root is possible.

With each such physically feasible root, a critical point can evidently
be associated. The way to have any appreciation of the behaviour of 
these critical points has already been outlined in Section~\ref{sec3}.
The first task would be to set up an autonomous dynamical system (in
terms of a mathematical parameter, $\tau$), which will be 
\begin{eqnarray}
\label{pseudynsys}
\frac{\mathrm d}{{\mathrm d}\tau}(v^2) &=&
2v^2\left[2c_{\mathrm s}^2-r\phi^{\prime}(r)\right] \nonumber \\
\frac{{\mathrm d}r}{{\mathrm d}\tau} &=& 
r \left(v^2-c_{\mathrm s}^2\right) . 
\end{eqnarray}
Subject to the perturbation scheme,  
$v^2 = v_{\mathrm{c}}^2 + \delta v^2$, $c_{\mathrm{s}}^2 =
c_{\mathrm{sc}}^2 + \delta c_{\mathrm{s}}^2$ and
$r = r_{\mathrm{c}} + \delta r$, equation~(\ref{pseudynsys}) will 
lead to a set of coupled autonomous linear equations in the perturbed 
quantities $\delta v^2$ and $\delta r$, with $\delta c_{\mathrm{s}}^2$ 
having first been expressed in terms of $\delta v^2$ and $\delta r$ 
from the continuity condition, as 
\begin{equation}
\label{varsound1}
\frac{\delta c_{\mathrm{s}}^2}{c_{\mathrm{sc}}^2} = -\frac{1}{2n}
\left[\frac{\delta v^2}{v_{\mathrm{c}}^2}
+ 4\frac{\delta r}{r_{\mathrm{c}}}\right] .
\end{equation}
The coupled linear dynamical system will be 
\begin{eqnarray}
\label{lindyn1}
\frac{\mathrm{d}}{\mathrm{d}\tau}(\delta v^2) &=&
- \frac{2c_{\mathrm{sc}}^2}{n}\,\delta v^2 - 2v_{\mathrm{c}}^2
{\mathcal D}\,\delta r \nonumber \\
\frac{\mathrm{d}}{\mathrm{d}\tau}(\delta r)
&=& r_{\mathrm{c}}\left(1+\frac{1}{2n}\right)\,\delta v^2
+ \frac{2c_{\mathrm{sc}}^2}{n}\,\delta r ,
\end{eqnarray}
with 
\begin{displaymath}
\label{dee}
{\mathcal D} = 
\frac{4c_{\mathrm{sc}}^2}{nr_{\mathrm{c}}} +
\phi^{\prime}(r_{\mathrm{c}}) + r_{\mathrm{c}}
\phi^{\prime \prime}(r_{\mathrm{c}}) . 
\end{displaymath} 

\begin{figure}
\begin{center}
\includegraphics[scale=0.33, angle=-90]{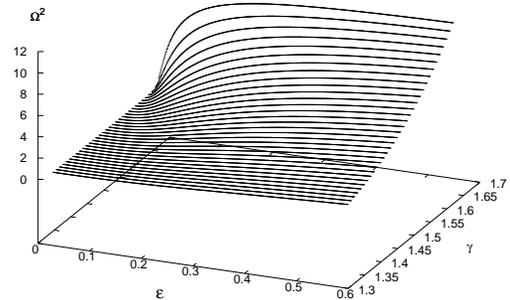}
\caption{\label{f3} \small{Dependence of $\Omega^2$ (always positive)
on the parameters $\mathcal E$ and $\gamma$ for the pseudo-Newtonian
potential, $\phi_1$. The saddle-type behaviour is maximum for intermediate
values of $\mathcal E$ and $\gamma$.}}
\end{center}
\end{figure}

\begin{figure}
\begin{center}
\includegraphics[scale=0.33, angle=-90]{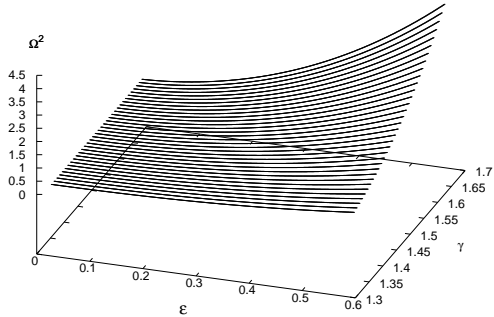}
\caption{\label{f4} \small{Dependence of $\Omega^2$ (always positive)
on the parameters $\mathcal E$ and $\gamma$ for the pseudo-Newtonian 
potential, $\phi_4$. The saddle-type behaviour has a monotonic growth 
with increasingly higher values of $\mathcal E$ and $\gamma$.}}
\end{center}
\end{figure}

From here it is an easy passage to deriving the eigenvalues of the
stability matrix associated with the critical points. With the use 
of solutions, $\delta v^2 \sim \exp(\Omega \tau)$ and 
$\delta r \sim \exp(\Omega \tau)$, in equations~(\ref{lindyn1}),
these eigenvalues will be derived as
\begin{equation}
\label{eigen1}
\Omega^2 = 
\frac{\left[r_{\mathrm{c}}\phi^{\prime}(r_{\mathrm{c}})\right]^2}{2}
\left[\left(3 - 5\gamma \right) - \left(\gamma +1\right)r_{\mathrm{c}}
\frac{\phi^{\prime \prime}(r_\mathrm{c})}
{\phi^{\prime}(r_\mathrm{c})}\right] ,
\end{equation}
from whose structure it can once again be claimed that the critical 
points can only be either saddle points or centre-type points. The
dependence of $\Omega^2$ on $\mathcal E$ and $\gamma$ has been separately 
shown in Figs.~\ref{f3} and~\ref{f4}, under the choice of $\phi_1$ and 
$\phi_4$, respectively. For $\phi_1$ there are two critical points, of
which only one is the physically relevant saddle point, while for $\phi_4$
there is only one critical point, which has to be found numerically. It
is always a saddle point. The two potentials, $\phi_1$ and $\phi_4$, 
have been chosen because they give a closer Newtonian approximation to 
fully general relativistic conditions, than the other two potentials,
$\phi_2$ and $\phi_3$~\citep{ds01}. From both the plots in 
Figs.~\ref{f3} and~\ref{f4}, it is quite evident that contrary to what 
it was for the fully general relativistic case, for pseudo-Schwarzschild 
flows, strongly transonic features (indicated by high positive values 
of $\Omega^2$) occur at much greater values of $\mathcal E$ and $\gamma$. 

\section{Concluding remarks}
\label{sec5}

In trying to mathematically understand the nature of the three 
critical points in the general relativistic flow, it has been shown
that there arises a situation, whereby, because of the fluctuating
nature of the middle critical point (sometimes a saddle point and
sometimes a centre-type point), two contiguous critical points
will be of the same kind. If these two particular points are both
saddle-type then as an exercise of mathematical interest 
(if not of any direct physical relevance), it will be patently
impossible to connect the two points by continuous solutions. However, 
going back to equation~(\ref{dvdr}), and subjecting it to a closer 
examination, a possible way of bypassing this difficulty could be 
found. It has been discussed already that the only critical conditions
selected from equation~(\ref{dvdr}) will be the ones on length 
scales greater than that of the event horizon. On the other hand, 
if some of the unphysical 
criteria ($v^2 = 1$ at $r=1$, or $v=0$ at $r=0$, etc.) for criticality
in the flow were to be taken into account, then some new critical 
points (of mathematical interest only) could be found in the global
phase portrait, even on unrealistic length scales. These possible 
critical points may then settle the difficulty which has arisen 
from the apparent existence of adjacent critical points of the same
nature. Solutions could then be connected from one region to the 
other through these ``hidden" critical points. 

Another problem with saddle points is that solutions passing through
them are notoriously sensitive to the fine tuning of the outer 
boundary condition of the flow. This is a standing 
problem with stationary flows and in consequence of this, it has been 
shown for the classical Bondi problem that transonicity could be 
achieved only if the evolution of the flow were to be followed through
time~\citep{rb02}. This is a relatively easy proposition in the 
Newtonian domain. When a flow is studied in the general relativistic
regime, the time-dependent evolution will require much greater mathematical
(both analytical and numerical) sophistication. Having made this point
it should also be a fair expectation that transonicity would continue
all the same to hold its primary position in spherically symmetric 
flows. 

\section*{Acknowledgements}

This research has made use of NASA's Astrophysics Data System. The 
authors express their indebtedness to J. K. Bhattacharjee, J. Ehlers,
A. D. Gangal, T. Naskar and Y. Shtanov for some useful comments.


\begin{thebibliography}{99}

\bibitem[\protect\citeauthoryear{Afshordi \& Paczy\'nski}{2003}]{ap03}
Afshordi, N., Paczy\'nski, B.,  2003, ApJ, 592, 354

\bibitem[\protect\citeauthoryear{Artemova et al.}{1996}]{abn96}
Artemova, I. V., Bj\"ornsson, G., Novikov, I. D.,  1996, ApJ, 461, 565 

\bibitem[\protect\citeauthoryear{Axford \& Newman}{1967}]{an67}
Axford, W. I., Newman, R. C.,  1967, ApJ, 147, 230

\bibitem[\protect\citeauthoryear{Balazs}{1972}]{bal72}
Balazs, N. L.,  1972, MNRAS, 160, 79

\bibitem[\protect\citeauthoryear{Begelman}{1978}]{beg78}
Begelman, M. C.,  1978, A\&A, 70, 53

\bibitem[\protect\citeauthoryear{Blumenthal \& Mathews}{1976}]{blum76}
Blumenthal, G. R., Mathews, W. G.,  1976, ApJ, 203, 714

\bibitem[\protect\citeauthoryear{Bohr et al.}{1993}]{bdp93}
Bohr, T., Dimon, P., Putkaradze, V.,  1993, Journal of Fluid 
Mechanics, 254, 635

\bibitem[\protect\citeauthoryear{Bonazzola et al.}{1987}]{bona87a}
Bonazzola, S., Falgarone, E., Heyvaerts, J., P\' {e}rault, M.,
Puget, J. L.,  1987, A\&A, 172, 293

\bibitem[\protect\citeauthoryear{Bonazzola et al.}{1992}]{bona92b}
Bonazzola, S., P\' {e}rault, M., Puget, J. L., Heyvaerts, J.,
Falgarone, E., Panis, J. F.,  1992, Journal of Fluid Mechanics, 245, 1

\bibitem[\protect\citeauthoryear{Bondi}{1952}]{bon52}
Bondi, H.,  1952, MNRAS, 112, 195

\bibitem[\protect\citeauthoryear{Brinkmann}{1980}]{bri80}
Brinkmann, W.,  1980, A\&A, 85, 146

\bibitem[\protect\citeauthoryear{Chakrabarti}{1990}]{skc90}
Chakrabarti, S. K.,  1990, Theory of Transonic Astrophysical
Flows, World Scientific, Singapore

\bibitem[\protect\citeauthoryear{Chakrabarti}{1996}]{skc96}
Chakrabarti, S. K.,  1996, Physics Reports, 266, 229

\bibitem[\protect\citeauthoryear{Chandrasekhar}{1939}]{sc39}
Chandrasekhar, S.,  1939, An Introduction to the Study of Stellar
Structure, The University of Chicago Press, Chicago

\bibitem[\protect\citeauthoryear{Chaudhury et al.}{2006}]{crd06}
Chaudhury, S., Ray, A. K., Das, T. K.,  2006, MNRAS, 373, 146

\bibitem[\protect\citeauthoryear{Cowie et al.}{1978}]{cos78}
Cowie, L. L., Ostriker, J. P., Stark, A. A.,  1978, ApJ, 226, 1041

\bibitem[\protect\citeauthoryear{Das}{1999}]{das99}
Das, T. K.,  1999, MNRAS, 308, 201

\bibitem[\protect\citeauthoryear{Das}{2000}]{das00}
Das, T. K.,  2000, MNRAS, 318, 294

\bibitem[\protect\citeauthoryear{Das}{2004}]{das04}
Das, T. K.,  2004, Classical and Quantum Gravity, 21, 5253

\bibitem[\protect\citeauthoryear{Das \& Sarkar}{2001}]{ds01}
Das, T. K., Sarkar, A.,  2001, A\&A, 374, 1150

\bibitem[\protect\citeauthoryear{Frank et al.}{2002}]{fkr02}
Frank, J., King, A., Raine, D.,  2002, Accretion Power in
Astrophysics, Cambridge University Press, Cambridge

\bibitem[\protect\citeauthoryear{Gaite}{2006}]{gai06}
Gaite, J.,  2006, A\&A, 449, 861

\bibitem[\protect\citeauthoryear{Garlick}{1979}]{gar79}
Garlick, A. R.,  1979, A\&A, 73, 171

\bibitem[\protect\citeauthoryear{Jordan \& Smith}{1999}]{js99}
Jordan, D. W., Smith, P.,  1999, Nonlinear Ordinary Differential
Equations, Oxford University Press, Oxford

\bibitem[\protect\citeauthoryear{Kazhdan \& Murzina}{1994}]{km94}
Kazhdan, Y. M., Murzina, M.,  1994, MNRAS, 270, 351

\bibitem[\protect\citeauthoryear{Kovalenko \& Eremin}{1998}]{ke98}
Kovalenko, I. G., Eremin, M. A.,  1998, MNRAS, 298, 861

\bibitem[\protect\citeauthoryear{Malec}{1999}]{malec99}
Malec, E.,  1999, Phys. Rev. D, 60, 104043

\bibitem[\protect\citeauthoryear{Markovic}{1995}]{mar95}
Markovic, D.,  1995, MNRAS, 277, 11

\bibitem[\protect\citeauthoryear{M\'esz\'aros}{1975}]{mes75}
M\'esz\'aros, P.,  1975, A\&A, 44, 59

\bibitem[\protect\citeauthoryear{M\'esz\'aros \& Silk}{1977}]{ms77}
M\'esz\'aros, P., Silk, J.,  1977, A\&A, 55, 289

\bibitem[\protect\citeauthoryear{Michel}{1972}]{mich72}
Michel, F. C.,  1972, Astrophys. Space Sci., 15, 153

\bibitem[\protect\citeauthoryear{Moncrief}{1980}]{monc80}
Moncrief, V.,  1980, ApJ, 235, 1038

\bibitem[\protect\citeauthoryear{Nowak \& Wagoner}{1991}]{nw91}
Nowak, A. M., Wagoner, R. V.,  1991, ApJ, 378, 656 

\bibitem[\protect\citeauthoryear{Paczy\'nski \& Wiita}{1980}]{pw80}
Paczy\'nski, B., Wiita P. J.,  1980, A\&A, 88, 23

\bibitem[\protect\citeauthoryear{Parker}{1958}]{par58}
Parker, E. N.,  1958, ApJ, 123, 664

\bibitem[\protect\citeauthoryear{Parker}{1966}]{par66}
Parker, E. N.,  1958, ApJ, 143, 32

\bibitem[\protect\citeauthoryear{Petterson et al.}{1980}]{pso80}
Petterson, J. A., Silk, J., Ostriker, J. P.,  1980, MNRAS, 191, 571

\bibitem[\protect\citeauthoryear{Ray}{2003}]{ray03}
Ray, A. K.,  2003, MNRAS, 344, 1085

\bibitem[\protect\citeauthoryear{Ray \& Bhattacharjee}{2002}]{rb02}
Ray, A. K., Bhattacharjee, J. K.,  2002, Phys. Rev. E, 66, 066303

\bibitem[\protect\citeauthoryear{Ray \& Bhattacharjee}{2005}]{rb05}
Ray, A. K., Bhattacharjee, J. K.,  2005, ApJ, 627, 368

\bibitem[\protect\citeauthoryear{Stellingwerf \& Buff}{1978}]{sb78}
Stellingwerf, R. F., Buff, J.,  1978, ApJ, 221, 661

\bibitem[\protect\citeauthoryear{Theuns \& David}{1992}]{td92}
Theuns, T., David, M.,  1992, ApJ, 384, 587

\bibitem[\protect\citeauthoryear{Titarchuk et al.}{1996}]{tmk96}
Titarchuk, L., Mastichiadis, A., Kylafis, N. D.,  1996, A\&A, 120, 171

\bibitem[\protect\citeauthoryear{Titarchuk et al.}{1997}]{tmk97}
Titarchuk, L., Mastichiadis, A., Kylafis, N. D.,  1997, ApJ, 487, 834

\bibitem[\protect\citeauthoryear{Toropin et al.}{1999}]{ttsrcl99}
Toropin, Yu. M., Toropina, O. D., Savelyev, V. V., Romanova, M. M.,
Chechetkin, V. M., Lovelace, R. V. E.,  1999, ApJ, 517, 906

\bibitem[\protect\citeauthoryear{Tsuribe et al.}{1995}]{tuf95}
Tsuribe, T., Umemura, M., Fukue, J.,  1995, PASJ, 47, 73

\bibitem[\protect\citeauthoryear{Vitello}{1984}]{vit84}
Vitello, P.,  1984, ApJ, 284, 394

\bibitem[\protect\citeauthoryear{Zampieri et al.}{1996}]{zmt96}
Zampieri, L., Miller, J. C., Turolla, R.,  1996, MNRAS, 281, 1183

\end{thebibliography}
\end{document}